\documentclass[
reprint,
superscriptaddress,
showpacs,
amsmath,amssymb,
aps,
]{revtex4-1}

\usepackage{graphicx}
\usepackage{dcolumn}
\usepackage{bm}
\usepackage{amsmath}
\usepackage{autobreak}
\usepackage{hyperref}
\usepackage[mathlines]{lineno}
\usepackage{natbib}
\usepackage{multirow}
\usepackage{mathtools}
\usepackage{color}
\usepackage{booktabs}
\usepackage{soul, color, xcolor}
\usepackage{appendix}
\usepackage{booktabs} 

\usepackage{bm}
\usepackage{array}
\usepackage{makecell}

\usepackage{float}

\begin{document}
	
\preprint{APS/123-QED}
\title{Calculation and comparison of sensitivities in $0\nu\beta\beta$ experiments based on key parameters}
\author{X.~Yu}
\affiliation{Key Laboratory of Particle and Radiation Imaging (Ministry of Education) and Department of Engineering Physics, Tsinghua University, Beijing 100084}
\author{L.~T.~Yang}\altaffiliation [Corresponding author: ]{yanglt@mail.tsinghua.edu.cn}
\affiliation{Key Laboratory of Particle and Radiation Imaging (Ministry of Education) and Department of Engineering Physics, Tsinghua University, Beijing 100084}
\author{Q.~Yue}\altaffiliation [Corresponding author: ]{yueq@mail.tsinghua.edu.cn}
\affiliation{Key Laboratory of Particle and Radiation Imaging (Ministry of Education) and Department of Engineering Physics, Tsinghua University, Beijing 100084}
\author{H.~Ma}
\affiliation{Key Laboratory of Particle and Radiation Imaging (Ministry of Education) and Department of Engineering Physics, Tsinghua University, Beijing 100084}
\author{H.~T.~Wong}
\affiliation{Institute of Physics, Academia Sinica, Taipei 11529}

\date{\today}

\begin{abstract}
Worldwide efforts are underway to detect neutrinoless double beta ($0\nu\beta\beta$) decay using experiments based on various technologies and target isotopes. Future experiments in this regard aim to exclude the inverted order (IO) condition or explore the normal order (NO) band. Consequently, comparing the sensitivities of proposed $0\nu\beta\beta$ decay experiments with promising prospects is essential. The current study adopts sensitivity metrics, including exclusion and discovery sensitivities, half-life sensitivities, and $m_{\beta\beta}$ sensitivities, to provide a comprehensive evaluation of ten typical promising experiments: LEGEND, CDEX, nEXO, XLZD, PandaX, NEXT, KamLAND-Zen, JUNO, SNO+, and CUPID, and highlight their unique features. Based on reported experimental parameters, the concept of a ``technical line'' is introduced to determine the location that each experiment may realize in the $\xi$ and $\lambda_{b}$ space, where $\xi$ represents the sensitive exposure per year, and $\lambda_{b}$ denotes the expected annual rate of background events. Half-life sensitivities for the selected experiments are calculated, some of them in multiple phases and others in conservative or aggressive condition. The results indicate that increasing the operation time is more beneficial for zero-background experiments, which also demonstrate greater competitiveness in discovery sensitivity. $m_{\beta\beta}$ sensitivities are presented as uncertainty bands arising from the nuclear matrix element uncertainties. Additionally, half-life and $m_{\beta\beta}$ sensitivities are estimated under ideal conditions also in the form of uncertainty bands, where only irreducible $2\nu\beta\beta$ and solar B-8 neutrino induced background remain.
\end{abstract}

\maketitle

\section{Introduction}\label{sec:1}
\subsection{Neutrinoless double beta decay}\label{sec:1a}
Neutrinoless double beta ($0\nu\beta\beta$) decay is a hypothetical physical process extending beyond the Standard Model, wherein two neutrons simultaneously transform into two protons and two electrons without releasing a neutrino: $(A,Z) \rightarrow (A,Z+2) + 2e^-$~\cite{status2019}. The observation of $0\nu\beta\beta$ decay is expected to provide the strongest experimental evidence supporting the Majorana nature of neutrinos and the violation of lepton number conservation~\cite{Laura2019}. Given its potential implications for new physics, extensive global research efforts have been dedicated toward the detection of $0\nu\beta\beta$ decay signals.

Although $0\nu\beta\beta$ signals have not yet been observed, researchers have consistently extended the lower limit of the $0\nu \beta \beta$ decay half-life through experiments utilizing diverse technologies and isotopes of interest~\cite{arnquist2023final,abe2023search,agostini2020final,armengaud2011new}. There are generally two paths of increasing the sensitivity of experiments, raising exposure and reducing background. Currently, the optimal lower limit of the $0\nu \beta \beta$ decay half-life has an order of magnitude of $10^{26}$ yr [90$\%$ confidence level (CL)]~\cite{arnquist2023final,abe2023search,agostini2020final}. To realize further improvements, ongoing and planned experiments aim to achieve a multiple-stage goal of reaching sensitivities of $10^{27}$ yr, $10^{28}$ yr, and even $10^{29}$ yr~\cite{abgrall2017large,abgrall2021legend,piepke2018sensitivity,albanese2021sno+,Zhao_2017,prospect_cjpl}.

In the pursuit of $0\nu\beta\beta$, there is going to be more attention to the irreducible background of $2\nu\beta\beta$. $2\nu\beta\beta$ signals cannot be distinguished from $0\nu\beta\beta$ signals because they have the same topology. In addition, elastic scatter between electron and solar B-8 neutrino also cannot be distinguished in most experiments except for nEXO~\cite{Adhikari_2022} and NEXT~\cite{NEXT_review}, which may use technique to detect Ba ions, the daughter nucleus of $2\nu\beta\beta$, in the future~\cite{agostini2023toward}. These irreducible backgrounds will contribute to a non-negligible percentage of total background for experiments with bad resolutions, especially future large-scale liquid scintillators, such as JUNO and SNO+~\cite{albanese2021sno+,Zhao_2017}. On the other hand, they are still negligible in experiments with high resolutions, such as High Purity Germanium Detectors like LEGEND and CDEX~\cite{abgrall2021legend,CDEX300,prospect_cjpl}. Therefore, the issue of irreducible backgrounds and the upgrade of resolution is essential for large-scale liquid scintillators, detailed discussion in Sec.~\ref{sec:4b}.

\begin{figure}[!htbp]
 	\includegraphics[width=\linewidth]{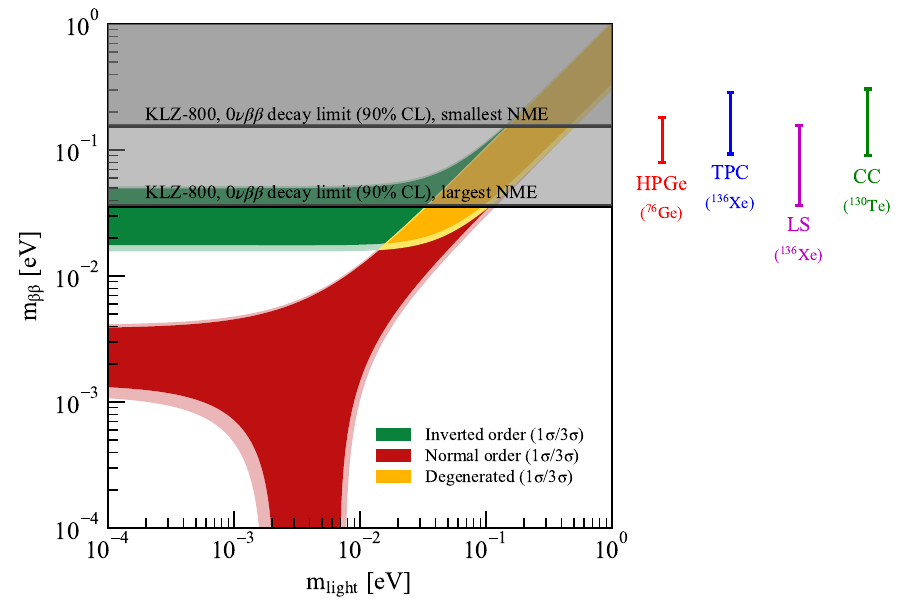}
 	\caption{Possible parameter spaces for the inverted order (IO) and normal order (NO) bands, along with their overlapping part. The best upper limit of $m_{\beta\beta}$ is acquired from KLZ-800, belonging to the KamLAND-Zen Project~\cite{abe2023search}. The best results of the four primary technologies come from GERDA~\cite{agostini2020final}, EXO-200~\cite{EXO-200}, KLZ-800~\cite{abe2023search}, and CUORE~\cite{CUORE}, which are demonstrated on the right.}
 	\label{fig::region}
\end{figure}

Notably, the lower limits of the $0\nu\beta\beta$ decay half-life differ across target isotopes owing to variations in nuclear parameters. Generally, if the mass mechanism induces $0\nu\beta\beta$ decay, the effective Majorana mass ($m_{\beta\beta}$) of the neutrino drives the process and plays an identical role across all isotopes of interest (as further detailed in Sec.~\ref{sec:2a}). Combining results from neutrino oscillation experiments and cosmological observations can help constrain the upper limit of $m_{\beta\beta}$. Recent $0\nu\beta\beta$ decay experiments have begun probing the inverted order (IO) band, as depicted in Fig.~\ref{fig::region}. Future experimental efforts aim to further lower the upper limit of $m_{\beta\beta}$ and ultimately exclude the entire IO band.

\subsection{Research status}\label{sec:1b}
Currently, various detector technologies are employed in $0\nu\beta\beta$ decay experiments worldwide. The four primary ones include high-purity germanium (HPGe) detectors, time projection chambers (TPCs), liquid scintillators (LSs), and cryogenic calorimeters (CCs).

HPGe detectors are semiconductor devices configured as string arrays of single crystals, with $^{76}\rm{Ge}$ serving as the target isotope. These detectors discriminate the energies and modes of physical events based on signals generated by ionized electrons and holes. HPGe detectors are known for their high efficiency, excellent energy resolution, and low background noise. Currently, the GERDA experiment is known to yield the best $0\nu \beta \beta$ decay outcome for $^{76}\rm{Ge}$: $T_{1/2}^{\rm{Ge}}>1.8\times 10^{26}~\rm{yr}$ (90$\%$ CL) and $m_{\beta\beta}<(79-180)~\rm{meV}$ (90$\%$ CL)~\cite{agostini2020final}. Notably, HPGe is the only detection technology that has successfully achieved zero-background operation. In the LEGEND-200 experiment~\cite{LEGEND200_TAUP2023}---a collaboration between GERDA and MJD---over 142 kg of HPGe detectors have been installed, with ongoing data collection. Plans for future heavier detectors with masses of 200 kg and 1000 kg are aimed at achieving sensitivities of $10^{27}$ yr and $10^{28}$ yr for the half-life of $^{76}\rm{Ge}$ $0\nu\beta\beta$ decay, respectively~\cite{abgrall2017large,abgrall2021legend}. Additionally, for the CDEX experiment at the China Jinping Underground Laboratory, a 300 kg HPGe detector is being designed~\cite{CDEX300}, with future upgrades targeting one-ton and ten-ton scales~\cite{prospect_cjpl}.

In contrast, TPCs mainly utilize $^{136}\rm{Xe}$ as their target isotope. These systems detect ionized electrons and ions drifting along an electric field in a liquid or gaseous xenon medium, and at the same time scintillation light generated during these events for distinguishing $0\nu \beta \beta$ decay events. Experiments such as EXO, NEXT, PandaX, and LZ employ the TPC technology, with different designs either under development or actively collecting data. Among these experiments, the EXO-200 experiment currently yields the best lower limit for $^{136}\rm{Xe}$ $0\nu \beta \beta$ decay: $T_{1/2}^{\rm{Xe}}>3.5\times 10^{25}~\rm{yr}$ (90$\%$ CL) and $m_{\beta\beta}<(93-286)~\rm{meV}$ (90$\%$ CL)~\cite{EXO-200}. nEXO~\cite{piepke2018sensitivity}, the next phase of the EXO-200 experiment, with a projected five-ton liquid enriched xenon detector, is expected to achieve a projected sensitivity of nearly $10^{28}$ years. In addition, the search for $0\nu\beta\beta$ is also one of the multiple purposes of large-scaled TPC experiments using natural Xe, such as XLZD~\cite{XLZD2024} and PandaX~\cite{PandaX2025}. A two-phase XLZD experiment utilizing 60 and 80 tons of liquid natural Xe, is also expected to expand the half-life sensitivity to $10^{28}$ yr. PandaX-xT, with 47 tons of natural Xe, is expected to reach a half-life sensitivity of $6.9 \times 10^{27}$ yr. TPCs are advantageous owing to their large scale and low background levels. However, only $0\nu \beta \beta$ decay events occurring within the fiducial volume of the inner xenon region are recorded, as the outer xenon medium acts as a background shield, resulting in relatively low detection efficiency. Additionally, the energy resolution of TPCs is inferior to that of crystal-based detectors such as HPGe and CCs.

LS technology represents a successful departure from the ``source = detector'' paradigm. In this approach, a target isotope is dissolved in the scintillator medium, and photomultiplier tubes surrounding the scintillator collect photons generated by scintillation events. This setup allows for the reconstruction of the energy, topology, and position of the detected events. Two primary target isotopes are used in LS experiments: $^{136}\rm{Xe}$ and $^{130}\rm{Te}$. So far, the KLZ-800 experiment, belonging to the KamLAND-Zen group, is known to offer the best lower limit for $0\nu\beta\beta$ decay: $T_{1/2}^{\rm{Xe}}>2.3\times 10^{26}~\rm{yr}$ (90$\%$ CL) and $m_{\beta\beta}<(36-156)~\rm{meV}$ (90$\%$ CL)~\cite{abe2023search}. Upcoming LS-based experiments include KL2Z~\cite{abe2023search}, an upgraded version of KamLAND-Zen focused on $^{136}\rm{Xe}$, and SNO+~\cite{albanese2021sno+}, an upgraded version of SNO using $^{130}\rm{Te}$. Both these experiments are designed for $0\nu\beta\beta$ decay detection with a sensitivity of $10^{27}$ yr. Additionally, JUNO, with its large-scale design, demonstrates the potential for $0\nu\beta\beta$ decay detection at a sensitivity of $10^{28}$ yr after determining the neutrino mass ordering~\cite{Zhao_2017}. While LSs are easier to operate than crystal detectors, their sensitivity is constrained by relatively high background levels, limited energy resolution, and the scalability of the scintillators.

CC is another class of crystal detectors, besides HPGe detectors,that utilize heat sensors to measure the temperature changes caused by one energy deposition event. These detectors utilize target nuclides such as $^{130}\rm{Te}$ and $^{100}\rm{Mo}$ in crystal form. So far, the CUORE experiment is known to offer the best lower limit for $^{130}\rm{Te}$ $0\nu \beta \beta$ decay using CCs: $T_{1/2}^{\rm{Te}}>2.2\times10^{25}~\rm{yr}$ (90$\%$ CL) and $m_{\beta\beta}<(90-305)~\rm{meV}$ (90$\%$ CL)~\cite{CUORE}. Future developments in CC technology include incorporating scintillation light readout for particle discrimination, which can further reduce background noise. CUPID, a next-generation experiment building on the CUORE infrastructure, aims to reach a sensitivity of $10^{27}$ yr and more for the $0\nu\beta\beta$ of $^{100}\rm{Mo}$~\cite{CUPID2022}. CCs offer advantages such as high energy resolution, efficiency, and low background levels owing to their crystal structure.

In conclusion, the four detection technologies discussed above---HPGe, TPCs, LSs, and CCs---exhibit distinct advantages across key parameters, including scalability, detection efficiency, energy resolution, and background suppression.

\subsection{Research objective}\label{sec:1c}
With the continuous emergence of new detectors based on diverse technologies, exploring the prospects and challenges of the four main detection technologies is essential. Such an analysis would offer valuable guidance for new research groups seeking appropriate detector technologies and serve as a reference for researchers in related fields.

Current research on the sensitivity estimation of $0\nu\beta\beta$ experiments can be divided into two types. One focuses on a certain experiment~\cite{piepke2018sensitivity,Adhikari_2022,XLZD2024} or a certain type of experiments~\cite{Compare2024}, usually containing physical analysis and simulation of multiple circumstances, which is a detailed type of research lack of generality. The other one, often occurring in review articles~\cite{agostini2023toward,status2019}, summarizes the sensitivities of various experiments for reference. However, little literature of this kind~\cite{agostini2023toward,Jason2017} gives detailed deduction of sensitivities from experimental parameters as well as comparison of different metrics of sensitivities.

To fill the gap mentioned above, this work aims to construct a framework of calculation from experimental parameters of typical experiments with different features to various metrics of sensitivities for comprehensive comparison. Such a framework is expected to be applicable to future proposed experiments.

There are two research goals in this study: (1) estimating the $T_{1/2}^{0\nu}$ and $m_{\beta\beta}$ sensitivities of representative promising experiments under different sensitivity metrics, covering both exclusion and discovery limits, and (2) giving analysis to an ideal condition when there is only $2\nu\beta\beta$ and elastic scattering (ES) induced by solar B-8 neutrinos contributing to the background, setting an upper limit for selected experiments in terms of theoretical physics. The parameters and formulas used in the sensitivity analysis are presented in Sec.~\ref{sec:2a}, while the sensitivity metrics are outlined in Sec.~\ref{sec:2b}. A new concept, the ``technical line,'' is introduced in Sec.~\ref{sec:3a} to distinguish the effects of exposure and performance per unit exposure. In Sec.~\ref{sec:3b}, the parameter space of each selected detector is visualized using a $\xi -\lambda _{b}$ diagram ($\xi$ denotes sensitive exposure per year, and $\lambda _{b}$ represents the expected annual background event). Based on the results detailed in Sec.~\ref{sec:3}, a multiple-phase estimation of the $T_{1/2}$ sensitivity of selected detectors is demonstrated in Sec.~\ref{sec:4a} while the ideal $T_{1/2}^{0\nu}$ sensitivity under irreducible background conditions is discussed in Sec.~\ref{sec:4b}. Similarly, a multiple-phase estimation of $m_{\beta\beta}$ sensitivity and the ideal $m_{\beta\beta}$ sensitivity is demonstrated in Sec.~\ref{sec:5}.

\section{Notations and formulas}\label{sec:2}
\subsection{Parameters and formulas}\label{sec:2a}
This subsection details parameters and formulas relevant to the physics and experimental design of $0\nu\beta\beta$ detectors. We adopt sensitivity metrics and notation from previous literature with minor modifications~\cite{agostini2023toward,singh2020exposure,Jason2017}.

When induced by the  ``mass mechanism," the $0\nu\beta\beta$ decay is driven by the effective Majorana mass $m_{\beta\beta}$. In this case, the decay half-life $T_{1/2}^{0\nu}$ is expressed as follows~\cite{singh2020exposure}:
\begin{equation}
	\begin{aligned}
		\label{Eq:halflife1}
		\frac{1}{T_{1/2}^{0\nu}}=G^{0\nu}g_{A}^{4}|M^{0\nu}|^{2}|\frac{m_{\beta\beta}}{m_{e}}|^{2}
	\end{aligned}
\end{equation}
where $G^{0\nu}$ represents the space phase factor~\cite{Phase2012}, $g_{A}$ denotes the effective axial vector coupling, $|M^{0\nu}|$ represents the nuclear matrix element, and $m_{e}$ denotes the electron mass. In the subsequent analysis, an ``unquenched'' free nucleon value of $g_{A}=1.27$ is used, while $|M^{0\nu}|$ ranges for the isotopes of interest are sourced from Ref.~\cite{agostini2023toward}. Given a value for $(T_{1/2}^{0\nu})^{-1}$, the corresponding $m_{\beta\beta}$ is determined by $\sqrt{G^{0\nu}|M^{0\nu}|^{2}}$ according to Eq.~(\ref{Eq:halflife1}). A larger $\sqrt{G^{0\nu}|M^{0\nu}|^{2}}$ value enables better limits for $m_{\beta\beta}$. Table~\ref{ta:parameter} lists the $\sqrt{G^{0\nu}|M^{0\nu}|^{2}}$ values of several target isotopes. These parameters are used in the sensitivity calculations described in Sec.~\ref{sec:5}. It is worth mentioning that spread of $|M^{0\nu}|$ for $^{100}\rm{Mo}$  is lower than other isotopes, but it is mostly likely due to lack of shell model estimation for $^{100}\rm{Mo}$, which often gives the lower limit for other isotopes. 

\begin{table}[!htbp]
	\begin{center}
		\renewcommand{\arraystretch}{1.5}
		\caption{$G^{0\nu}$, $|M^{0\nu}|$, and $\sqrt{G^{0\nu}|M^{0\nu}|^{2}}$ for various isotopes of interest, with $G^{0\nu}$ defined in units of $10^{-15}~\rm{yr}$~\cite{agostini2023toward,Phase2012}.}
		\label{ta:parameter}
		\begin{tabular}{p{1.2cm}<{\centering}p{1.6cm}<{\centering}p{1.8cm}<{\centering}p{1.8cm}<{\centering}p{1.5cm}<{\centering}}
			\hline
			\hline
			isotope& $G^{0\nu}$ \newline $(10^{-15}~\rm{yr})$& $|M^{0\nu}|$& $\sqrt{G^{0\nu}|M^{0\nu}|^{2}}$&$Q_{\beta\beta}$ \newline (keV)\\
			\hline
			$^{76}\rm{Ge}$&2.363&2.66--6.34&4.09--9.75&2039.1\\
			$^{136}\rm{Xe}$&14.58&1.11--4.77&4.24--18.21&2457.8\\
			$^{100}\rm{Mo}$&15.92&3.84--6.59&15.32--26.29&3034.4\\
			$^{130}\rm{Te}$&14.22&1.37--6.41&5.17--24.17&2527.5\\
			\hline
			\hline
		\end{tabular}
	\end{center}
\end{table}

Two key parameters characterize $0\nu\beta\beta$ decay detectors: sensitive exposure ($\xi$) and the expected background event rate per year ($\lambda_{b}$).

Sensitive exposure, $\xi$, is defined as the total mass of the target isotope ($m_{iso}$) multiplied by the detection efficiency ($\eta$). The efficiency $\eta$ is the product of several factors: active fraction of the target mass $\epsilon_{act}$, probability of the $0\nu\beta\beta$ decay energy being fully contained in the detector $\epsilon_{cont}$), multivariate analysis efficiency of tagging events in the sensitive volume $\epsilon_{mva}$, and efficiency of identifying a $0\nu\beta\beta$ event within the energy region of interest (ROI) $\epsilon_{ROI}$. Accordingly, $\xi$ is given as
\begin{equation}
	\begin{aligned}
		\label{eq:xi}
		\xi = m_{iso} \epsilon_{act} \epsilon_{cont} \epsilon_{mva} \epsilon_{ROI}=m_{iso}\eta 
	\end{aligned}
\end{equation} 
The expected background event rate $\lambda_{b}$ is defined as
\begin{equation}
	\begin{aligned}
		\label{lambda_b}
		\lambda_{b}=b\frac{m_{iso}A}{f_{enr}}\Delta E
	\end{aligned}
\end{equation}
where $m_{iso}$ means total amount of target isotope measured in the unit ``mol"; $b$ denotes the background per unit mass, time, and energy width ($\rm{\frac{counts}{keV\cdot kg\cdot yr }}$ or cpkky); $A$ represents the atomic mass of the target isotope; $f_{enr}$ signifies the enrichment fraction of the target isotope; and $\Delta E$ denotes the energy width of the ROI. Notably, in most experiments, ROIs are symmetrical, such as $[-2\sigma, 2\sigma]$, where $\sigma$ denotes the energy resolution. However, in some LS experiments, ROIs are skewed to minimize background noise.

\subsection{Sensitivity metrics}\label{sec:2b}
In the following sections, two standards for measuring sensitivity are used in parallel: exclusion sensitivity and discovery sensitivity. Among these, exclusion sensitivity refers to the expected number of signal events, $\lambda_{s}$, that an experiment has a 50\% probability of excluding at a 90\% CL. This parameter can be determined as follows:
\begin{equation}
	\label{Eq:exclusion sensitivity}
	\left\{
	\begin{aligned}
		&P(X\le x|\lambda_{b}T)\ge 50\%\\
		&P(X\ge x|\lambda_{b}T+\lambda _{s} )\ge 90\%,
	\end{aligned}
	\right.
\end{equation} 
where $P(X\le x|\lambda)$ represents the probability of observing $X$ signal events less than or equal to $x$ events when the expected signal number of events is $\lambda$, and $T$ denotes the operation time of the detector. For a given $\lambda_{b}$ and $T$, the minimum value of $x$ satisfying the first inequality is computed. This value is then used to determine the minimum $\lambda_{s}$ that satisfies the second inequality. Meanwhile, the discovery sensitivity refers to $\lambda_{s}$ for which an experiment has a 50\% probability of detecting an excess of events above the background at a 99.73\% CL.:
\begin{equation}
	\label{Eq:discovery sensitivity}
	\left\{
	\begin{aligned}
		&P(X\le x|\lambda_{b}T)\ge 99.73\%\\
		&P(X\ge x|\lambda_{b}T+\lambda _{s} )\ge 50\% 
	\end{aligned}
	\right.
\end{equation} 
Notably, the value of $\lambda_{s}(\lambda_{b}T)$ represents the number of $0\nu\beta\beta$ signal events corresponding to a certain sensitivity.

If the event distribution, $P(X|\lambda)$, follows a Poisson distribution, $\lambda_{s}$ does not increase with $\lambda_{b}T$ in intervals where $x$ remains constant. To address this, a new probability distribution interpolating the Poisson mass function is adopted. This distribution, which includes a normalized upper incomplete gamma function, aligns with the Poisson distribution when $x$ is an integer. Furthermore, in this new distribution, $x$ increases with the background, as detailed in Eq.~(\ref{Eq:distribution})~\cite{agostini2023toward}.
\begin{equation}
	\label{Eq:distribution}
	\begin{aligned}
		P(X\ge x|\lambda)=\frac{\Gamma(x+1,\lambda)}{\Gamma(x+1)}
	\end{aligned}
\end{equation} 

According to the abovementioned sensitivity definitions, the zero background condition occurs when $x=0$ satisfies the first inequality in Eq.~(\ref{Eq:exclusion sensitivity}) and Eq.~(\ref{Eq:discovery sensitivity}), where a single signal denotes the excess from the expected background. In this case, sensitivity does not improve with further reductions in background, resulting in a constant $\lambda_{s}$. Specifically, for exclusion sensitivity, the zero background condition occurs when $\lambda_{b}T<0.69$. Meanwhile, for discovery sensitivity, it occurs when $\lambda_{b}T<0.0027$. The second inequality in the exclusion and discovery sensitivity equations is satisfied when $\lambda_{s}=1.61$ and $\lambda_{s}=0.69$, respectively. To date, no experiments have been able to achieve the zero background condition for discovery sensitivity. However, experiments such as LEGEND-1000 have the potential to reach the zero background condition for exclusion sensitivity over a ten-year operational period. 

Given the operational time $T$ and the defined sensitivity rules $H$ (exclusion or discovery), the half-life sensitivity can be expressed as a function of $\varepsilon$ and $\lambda_{b}$, considering the ultralow detection rate of $0\nu\beta\beta$ decays,
\begin{equation}
 	\label{Eq:halflife2}
 	\begin{aligned}
 	T_{1/2}^{0\nu }=F(\xi ,\lambda_{b}  )=\frac{\rm{ln}2N_{A}\xi T  }{\lambda _{s}(\lambda _{b}T,H) }
 	\end{aligned} 	
 \end{equation}
According to Eq.~(\ref{Eq:halflife2}), the location of an experiment can be represented in a $\xi-\lambda_{b}$ diagram once its parameters are determined. Furthermore, if the expression on the left side of Eq.~(\ref{Eq:halflife2}) is set to a fixed value, a line called an equivalent line can be drawn on the $\xi-\lambda_{b}$ diagram. Points along this line correspond to experiments with the same half-life sensitivity, while points above this line correspond to experiments with higher sensitivity than the specified equivalent line. It is worth noting that the slope of the equivalent line approaches to 1/2 when $\lambda_{b}$ is large. Since $T_{1/2}^{0\nu}\propto \sqrt{\frac{MT}{b\Delta E}}$~\cite{agostini2023toward}, we can get the relation $\xi \propto \sqrt{\lambda_{b}}$ by replacing the parameters with Eq.~(\ref{eq:xi}) and Eq.~(\ref{lambda_b}) when $T_{1/2}^{0\nu}$ is fixed.

\begin{figure}[!htbp]
 	\includegraphics[width=\linewidth]{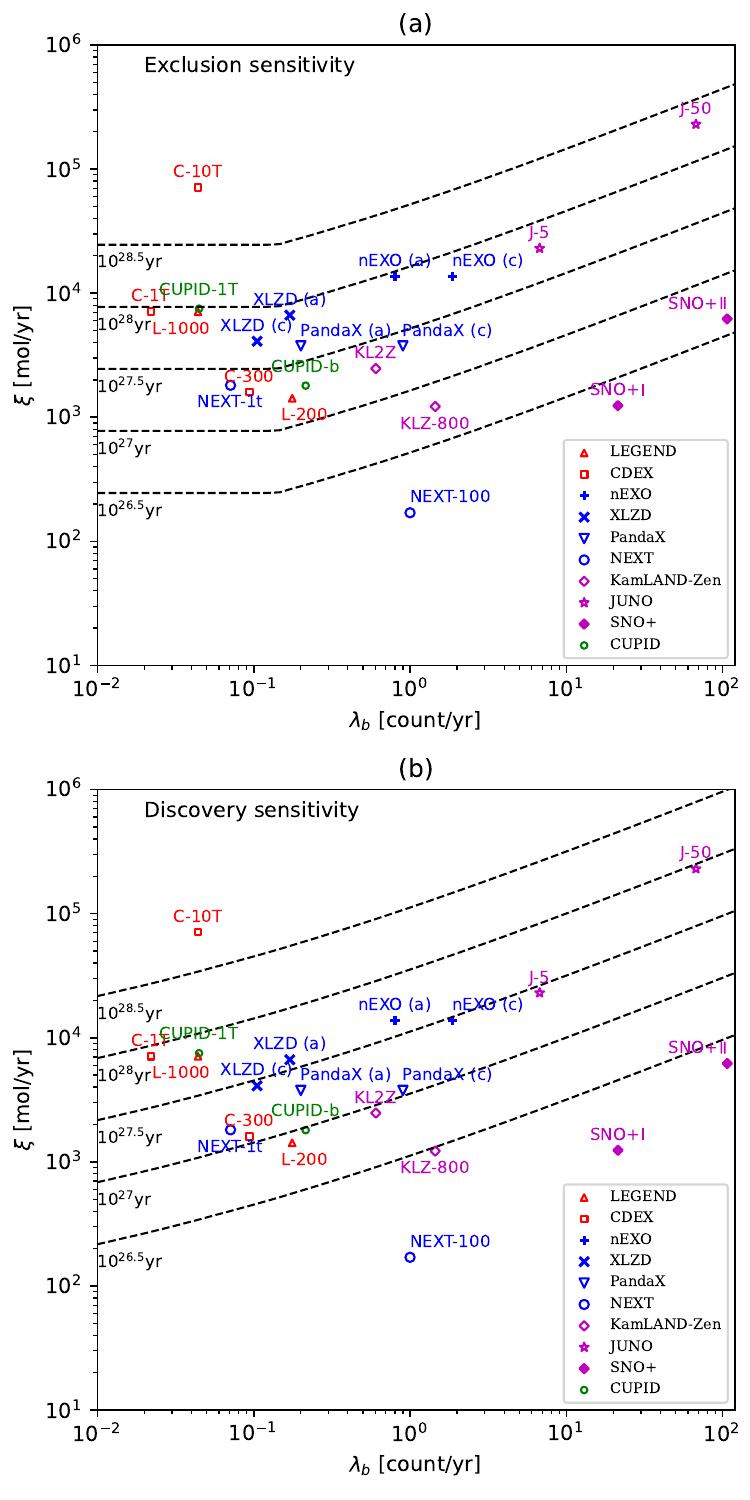}
 	\caption{Sensitive exposure $\xi$ and the expected background event rate per year $\lambda_{b}$ for major experiments. Locations of LEGEND, CDEX, nEXO, XLZD, PandaX, NEXT, KamLAND-Zen, JUNO, SNO+, and CUPID at different phases are calculated independently, with detailed information in Sec.~\ref{sec:3b}. The dashed lines represent the sets of vertices achieving specific values of (a) exclusion sensitivity and (b) discovery sensitivity for $T=5~\rm{yr}$. For $\lambda_{b}T<0.69$, the zero background condition of exclusion sensitivity is satisfied, and the equivalent line becomes parallel to the $\lambda_{b}$ axis. Red, blue, purple, and green scatters correspond to the HPGe, TPC, LS, and CC detectors, respectively.}
 	\label{fig:diagram}
\end{figure}

The locations of recent and future experiments in the $\xi-\lambda_{b}$ diagram are presented in Fig.~\ref{fig:diagram}. 

\section{Estimation method}\label{sec:3}
\subsection{Technical line}\label{sec:3a}
The sensitivity of an experiment can be improved through two primary approaches: increasing the exposure or enhancing the detector's performance at a fixed exposure. To increase exposure or expand the experiment's scale, researchers may either add more target isotope to the existing detector system or construct a larger one to accommodate greater exposure. Performance upgrades typically focus on reducing the background $b$, increasing the efficiency $\eta$, and improving the energy resolution $\sigma$. These two approaches are referred to as the ``scale'' path and ``performance'' path, respectively.

The technical line is introduced to distinguish the effects of the ``scale'' path and the ``performance'' path. By combining Eq.~(\ref{eq:xi}) with Eq.~(\ref{lambda_b}), a linear relationship between $\xi$ and $\lambda_{b}$ in logarithmic coordinates can be expressed as follows. The slope of this line is 1, while its intercept is $k$ (in unit of $\rm{\frac{mol}{events}}$), also known as the technical parameter.
\begin{equation}
	\label{Eq:technical line}
	\left\{
	\begin{aligned}
		&\rm{ln}\xi=\rm{ln}\lambda_{b}+\rm{ln}k\\
		&k=\frac{\eta f_{enr}}{bA\Delta E}
	\end{aligned}
	\right.
\end{equation}

\begin{figure}
	\includegraphics[width=\linewidth]{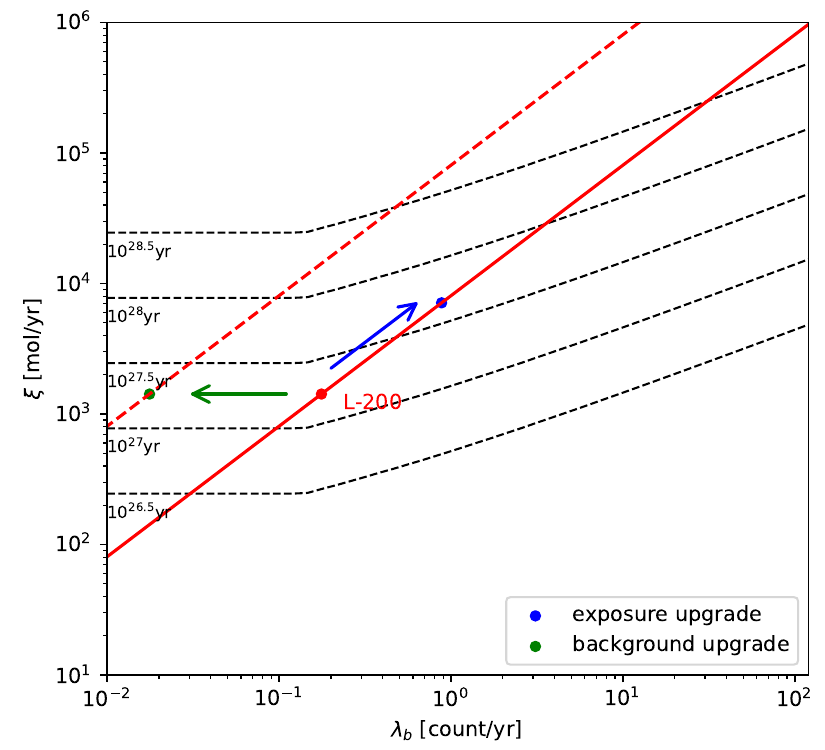}
	\caption{Location of LEGEND-200 in $\xi-\lambda_{b}$ diagram  when there is an upgrade in exposure or background level is shown. The solid red line corresponds to $k=8070$, technical parameter of LEGEND200. The dashed one corresponds to $k=80700$. (As is shown in Table~\ref{ta:LEGEND}) The blue scatter corresponds to LEGEND-200 exposure upgrade to the same of LEGEND-1000, while the green one corresponds to LEGEND-200 background upgrade to $b=1\times 10^{-5}$ cpkky.}
	\label{fig:technical_line}
\end{figure}

In the $(\xi,\lambda_{b})$ logarithmic diagram, technical lines are represented as parallel lines with a slope of 1 but different intercepts. For a specific detector, efforts to reduce background $b$, improve energy resolution $\sigma$, and increase efficiency $\eta$ result in increased $k$ values. At a fixed exposure, a larger $k$ value corresponds to better sensitivity. Conversely, increasing exposure with a fixed $k$ value is equivalent to moving a point upward along the same technical line in the $(\xi,\lambda_{b})$ logarithmic diagram. In summary, advancements along the ``performance'' path correspond to increasing $k$,  shifting the technical line upward. Meanwhile, advancements along the ``performance'' path correspond to moving upward along a fixed technical line in the diagram. Figure~\ref{fig:technical_line} gives a explicit demonstration of how the technical line demonstrates the upgrade of a certain experiment.
In some studies on TPC or LS experiments, the effective background in the fiducial volume, $b_{E}$, is used instead of $b$. The parameter $b_{E}$ represents the number of background events per keV width, per year, and per kilogram of the target isotope within the fiducial volume~\cite{agostini2023toward,Zhao_2017,piepke2018sensitivity,Adhikari_2022}. It is expressed as
\begin{equation}
	\label{Eq:b_E}
	\begin{aligned}
		b_{E}=\frac{\lambda_{b}}{Mf_{enr}\epsilon_{act}\Delta E}=\frac{\lambda_{b}}{m_{iso}A\epsilon_{act}\Delta E }
	\end{aligned}
\end{equation}
where $M$ denotes the total mass of the isotope. The technical parameter $k$ is reformulated for such experiments, allowing for straightforward calculation,
\begin{equation}
	\label{Eq:k_LS}
	\begin{aligned}
		k=\frac{\eta/\epsilon_{act}}{b_{E}\Delta E A}.
	\end{aligned}
\end{equation}

In conclusion, the introduction of $k$ separates the effect of different parameters on sensitivity and provides more insights into an experiment. A detailed discussion follows in the next subsection.

Notably, the sensitive background, $\mathcal{B} $, represents the number of expected background events in the sensitive volume normalized by $\xi$~\cite{agostini2023toward},
\begin{equation}
	\label{eq:B}
	\mathcal{B} =\lambda_{b}/\xi.
\end{equation}  
Interestingly, $\mathcal{B} $ is the reciprocal of $k$ but with a different physical interpretation. Previous studies considering $\mathcal{B} $ have exclusively focused on background issues without clearly connecting $\mathcal{B}$ to experimental specifications. In this study, $k$ is used instead, as it can be intuitively expressed as the intercept in the $(\xi,\lambda_{b})$ diagram. The value of $k$ is calculated under different experimental conditions to support this analysis.

\subsection{Experimental parameter demonstration} \label{sec:3b}
As discussed above in Eq.~(\ref{eq:xi}), Eq.~(\ref{Eq:technical line}), and Eq.~(\ref{Eq:k_LS}), the location of a given experiment in the $(\xi,\lambda_{b})$ logarithmic diagram can be calculated from experimental parameters, which is illustrated in Fig.~\ref{fig:diagram}. $\xi$, $k$, $\lambda_{b}$ and experimental parameters needed to calculate them as well as $T$ are either summarized in Table~\ref{ta:LEGEND} or Table~\ref{ta:nEXO} according to the expression of the background index. A detailed analysis can be found in the Appendix.

\begin{table*}[!htbp]
	\begin{center}
		\renewcommand{\arraystretch}{1.5}
		\caption{Specific parameters of LEGEND, CDEX, and CUPID required to estimate sensitivity for two phases each. EnGe means enriched germanium. The value of $\sigma/Q_{\beta\beta}$ is also presented for each experiment in the $\sigma$ column. The $\eta$ value for CUPID (in italics) is calculated inversely from estimated ten-year discovery sensitivity of CUPID baseline in the literature. The values of $\xi$ and $k$ are obtained from Eq.~(\ref{eq:xi}) and Eq.~(\ref{Eq:technical line}), respectively.}
		\label{ta:LEGEND}
		\begin{tabular}{p{3.1cm}<{\centering}p{1.0cm}<{\centering}p{1.6cm}<{\centering}p{1.4cm}<{\centering}p{1.0cm}<{\centering}p{1.8cm}<{\centering}p{1.8cm}<{\centering}p{1.1cm}<{\centering}p{1.4cm}<{\centering}p{1.6cm}<{\centering}p{1.0cm}<{\centering}}
			\hline
			\hline
			Phase&Setup&$m_{iso}$ ($\rm{mol}$)
			&$f_{enr}$ (\%)
			& $\eta$ (\%) 
			& $\sigma$ (keV) 
			&$b$ ($\frac{\rm{events}}{\rm{keV}\cdot\rm{kg}\cdot\rm{yr}}$)
			& $\xi$ $(\frac{\rm{mol}}{\rm{yr}})$
			& $k$ $(\frac{\rm{mol}}{\rm{events}})$
			& $\lambda_{b}$ $(\frac{\rm{events}}{\rm{yr}})$
			& $T$ $(\rm{yr})$\\
			\hline
			LEGEND-200~\cite{abgrall2021legend,LEGEND200_TAUP2023}&EnGe&2370&90&60&1.1 (0.054\%)&$2\times 10^{-4}$&1420&8070&0.176&5\\
			LEGEND-1000~\cite{abgrall2021legend,LEGEND1000_TAUP2023}&EnGe&11800&90&60&1.1 (0.054\%)&$1\times 10^{-5}$&7080&161000&0.044&10\\
			CDEX-300~\cite{CDEX300}&EnGe&2670&90&60&1.1 (0.054\%)&$9.5\times 10^{-5}$&1600&17000&0.094&5\\
			CDEX-1T~\cite{prospect_cjpl}&EnGe&11800&90&60&1.1 (0.054\%)&$5\times 10^{-6}$&7100&323000&0.022&10\\
			CDEX-10T~\cite{prospect_cjpl}&EnGe&118000&90&60&1.1 (0.054\%)&$1\times 10^{-6}$&71000&1615000&0.044&10\\
			CUPID baseline~\cite{CUPID2022}&$\rm{Li}_{2}\rm{Mn}\rm{O}_{4}$&2400&95&\emph{75}&2.1 (0.069\%)&$10^{-4}$&1800&8380&0.215&5 (10)\\
			CUPID-1T~\cite{CUPID2022}&$\rm{Li}_{2}\rm{Mn}\rm{O}_{4}$&10000&95&\emph{75}&2.1 (0.069\%)&$5 \times 10^{-6}$&7500&168000&0.045&10\\
			\hline
			\hline
		\end{tabular}
	\end{center}
\end{table*}

\begin{table*}[!htbp]
	\begin{center}
		\renewcommand{\arraystretch}{1.5}
		\caption{Specific parameters of NEXT, KamLAND-Zen, JUNO and SNO+ required to estimate sensitivity for two phases while parameters of nEXO, XLZD and PandaX-XT are, respectively, in conservative (c) and aggressive (a) conditions. EnXe means enriched xenon, NaXe and NaTe means natural xenon and tellurium. The value of $\sigma/Q_{\beta\beta}$ is also presented for each experiment in the $\sigma$ column. The $\eta$ value for nEXO (in italics) is inversely calculated from the estimated ten-year exclusion sensitivity in the literature. The values of $\xi$ and $k$ are obtained from Eq.~(\ref{eq:xi}) and Eq.~(\ref{Eq:k_LS}), respectively.}
		\label{ta:nEXO}
		\begin{tabular}{p{3.1cm}<{\centering}p{1.0cm}<{\centering}p{1.6cm}<{\centering}p{1.4cm}<{\centering}p{1.9cm}<{\centering}p{1.7cm}<{\centering}p{1.0cm}<{\centering}p{1.5cm}<{\centering}p{1.5cm}<{\centering}p{1.6cm}<{\centering}p{1.0cm}<{\centering}}
			\hline
			\hline
			Phase&Setup&$m_{iso}$ ($\rm{mol}$)
			& $\epsilon_{act}$ (\%)
			& $b_{E}$ ($\frac{\rm{events}}{\rm{keV}\cdot\rm{kg}\cdot\rm{yr}}$)
			& $\sigma$ ($\rm{keV}$)
			& $\eta$ (\%)
			& $\xi$ $(\frac{\rm{mol}}{\rm{yr}})$
			& $k$ ($\frac{\rm{mol}}{\rm{events}}$)
			& $\lambda_{b}$ $(\frac{\rm{events}}{\rm{yr}})$
	     	& $T$ $(\rm{yr})$\\
			\hline
			nEXO (c)~\cite{piepke2018sensitivity,Adhikari_2022}&EnXe&31800&68&$5.95\times 10^{-6}$&25 (1\%)&\emph{43}&13700&7360&1.86&10\\
			nEXO (a)~\cite{piepke2018sensitivity,Adhikari_2022}&EnXe&31800&68&$3.4\times 10^{-6}$&20 (0.81\%)&\emph{43}&13700&17100&0.801&10\\
			XLZD (c)~\cite{XLZD2024}&NaXe&39300&18&$3.4\times 10^{-6}$&16 (0.65\%)&10.4&4100&39000&0.105&10\\
			XLZD (a)~\cite{XLZD2024}&NaXe&52400&22&$3.4\times 10^{-6}$&16 (0.65\%)&12.7&6650&39000&0.171&10\\
			PandaX-xT (c)~\cite{PandaX2025}&NaXe&30800&18&$2.4\times 10^{-5}$&25 (1\%)&12.2&3760&4180&0.9&10\\
			PandaX-xT (a)~\cite{PandaX2025}&NaXe&30800&18&$5.35\times 10^{-6}$&25 (1\%)&12.2&3760&18800&0.2&10\\
			NEXT-100~\cite{NEXT_review,agostini2023toward}&EnXe&660&88&$4.4\times 10^{-4}$&10 (0.41\%)&26&170&170&1&5\\
			NEXT-1t~\cite{NEXT_review,agostini2023toward}&EnXe&7340&95&$4.4\times 10^{-6}$&5.2 (0.21\%)&24.6&1810&25500&0.071&10\\
			KLZ-800~\cite{abe2023search,agostini2023toward}&EnXe&5000&58&$2.5\times 10^{-5}$&105 (4.3\%)&24.4&1220&842&1.45&5\\
			KL2Z~\cite{Shirai_2017,agostini2023toward}&EnXe&7350&80&$1.1 \times 10^{-5}$&49 (2\%)&33.6&2470&4090&0.6&10\\
			JUNO (5 tons)~\cite{Zhao_2017}&EnXe&66300&45&$1.2 \times 10^{-5}$&47 (1.9\%)&34.7&23000&3410&6.74&5\\
			JUNO (50 tons)~\cite{Zhao_2017}&EnXe&545000&67&$1.2 \times 10^{-5}$&47 (1.9\%)&42.2&230000&3410&67.4&10\\
			SNO+\uppercase\expandafter{\romannumeral1}~\cite{SNO2016,SNO2020}&NaTe&10000&20&$3.57\times 10^{-4}$&115 (4.55\%)&12.4&1240&58&21.4&5\\
			SNO+\uppercase\expandafter{\romannumeral2}~\cite{SNO2016,SNO2020}&NaTe&50000&20&$3.57\times 10^{-4}$&115 (4.55\%)&12.4&6200&58&106.9&10\\
			\hline
			\hline
		\end{tabular}
	\end{center}
\end{table*}

Based on detector parameters sourced from the literature, we can calculate the value of the technical parameter $k$. In this study, ten typical experiments with strong potential are analyzed, covering all four major detector technologies and four major target isotopes listed in Table~\ref{ta:parameter}: LEGEND, CDEX, nEXO, XLZD, NEXT, KamLAND-Zen, JUNO, SNO+, and CUPID. Most of the experiments have a two-phase plan while CDEX has a three-phase plan. All the phases are analyzed separately in this subsection. On the other hand, the two TPC experiments, nEXO and XLZD, only have one phase. Therefore in the analysis they are divided by two possible conditions, conservative and aggressive, each with different experimental parameters. The experimental parameters are summarized in Table~\ref{ta:LEGEND} and Table~\ref{ta:nEXO}.

\section{Estimation of half-life sensitivity}\label{sec:4}
\subsection{Realistic half-life estimation}\label{sec:4a}
This section deals with the estimation of the exclusion and discovery of half-life sensitivities of the selected experiments. These calculations are based on Eq.~(\ref{Eq:halflife2}), using the values of $\xi$ and $k$ detailed in Table~\ref{ta:LEGEND} and Table~\ref{ta:nEXO}. Furthermore, $\lambda_s$ is either calculated for exclusion sensitivity or discovery sensitivity. The exclusion and discovery sensitivities for experiments in their early phases when $T=5~\rm{yr}$ are summarized in Table~\ref{ta:5yr_half-life}. Meanwhile, the exclusion and discovery sensitivities for later phases when $T\equiv 10~\rm{yr}$ and the exclusion sensitivity when $T\equiv 5~\rm{yr}$ are summarized in Table~\ref{ta:10yr_half-life}. A ten-year operation time is considered only for later phases. In contrast, early-phase experiments often undergo ongoing upgrades, making extended stable operation unlikely. The estimation of nEXO, XLZD and PandaX at later phases is based on an aggressive situation while early phases of the two experiments are not defined. Results are demonstrated in Table~\ref{ta:5yr_half-life} and Table~\ref{ta:10yr_half-life}.

\begin{table}[!tbp]
	\begin{center}
		\renewcommand{\arraystretch}{1.5}
		\caption{Exclusion and discovery half-life sensitivities of early phases of the selected experiments at $T=5~\rm{yr}$. Early phase is not defined for nEXO and XLZD. Lattice is in bold form if it is in zero background condition.}
		\label{ta:5yr_half-life}
		\begin{tabular}{p{2.3cm}<{\centering}p{1.0cm}<{\centering}p{2.2cm}<{\centering}p{2.2cm}<{\centering}}
			\hline
			\hline
			Experiment&Isotope&Exclusion $(\rm{yr})$ &Discovery $(\rm{yr})$\\
			\hline
			LEGEND-200&$^{76}\rm{Ge}$&$1.67\times 10^{27}$&$8.05\times 10^{26}$\\
			CDEX-300&$^{76}\rm{Ge}$&\bm{$2.06\times 10^{27}$}&$1.15\times 10^{27}$\\
			NEXT-100&$^{136}\rm{Xe}$&$1.04\times 10^{26}$&$4.81\times 10^{25}$\\
			KLZ-800&$^{136}\rm{Xe}$&$6.35\times 10^{26}$&$2.93\times 10^{26}$\\
			JUNO (5 tons)&$^{136}\rm{Xe}$&$6\times 10^{27}$&$2.76\times 10^{27}$\\
			SNO+\uppercase\expandafter{\romannumeral1}&$^{130}\rm{Te}$&$1.88\times 10^{26}$&$8.62\times 10^{25}$\\
			CUPID baseline&$^{100}\rm{Mo}$&$1.98\times 10^{27}$&$9.5\times 10^{26}$\\
			\hline
			\hline			
		\end{tabular}		
	\end{center}
\end{table}

\begin{table*}[!htbp]
	\begin{center}
		\renewcommand{\arraystretch}{1.5}
		\caption{Exclusion and discovery half-life sensitivities of later stages of the selected experiments at $T=10~\rm{yr}$ and exclusion sensitivity at $T=5~\rm{yr}$. The ratios of ten-yr and five-yr exclusion sensitivity are shown in the last column. Both nEXO and XLZD are in their aggressive condition. Lattice is in bold form if it is in zero background condition.}
		\label{ta:10yr_half-life}
		\begin{tabular}{p{2.5cm}<{\centering}p{1.2cm}<{\centering}p{3.2cm}<{\centering}p{3.2cm}<{\centering}p{3.2cm}<{\centering}p{1.2cm}<{\centering}}
			\hline
			\hline
			Experiment
			&Isotope
			&Five-yr exclusion $(\rm{yr})$
			&Ten-yr exclusion $(\rm{yr})$
			&Discovery $(\rm{yr})$&Ratio\\
			\hline
			LEGEND-1000&$^{76}\rm{Ge}$&\bm{$9.12\times 10^{27}$}&\bm{$1.82\times 10^{28}$}&$1.04\times 10^{28}$&2.0\\
			CDEX-1T&$^{76}\rm{Ge}$&\bm{$9.15 \times 10^{27}$}&\bm{$1.83\times 10^{28}$}&$1.32\times 10^{28}$&2.0\\
			CDEX-10T&$^{76}\rm{Ge}$&\bm{$9.15 \times 10^{28}$}&\bm{$1.83\times 10^{29}$}&$1.04\times 10^{29}$&2.0\\
			nEXO (a)&$^{136}\rm{Xe}$&$9.14 \times 10^{27}$&$1.36\times 10^{28}$&$6.3\times 10^{27}$&1.49\\
			XLZD (a)&$^{136}\rm{Xe}$&$7.97 \times 10^{27}$&$1.24\times 10^{28}$&$5.86\times 10^{27}$&1.56\\
			PandaX-xT (a)&$^{136}\rm{Xe}$&$4.26 \times 10^{27}$&$6.59 \times 10^{27}$&$3.11 \times 10^{27}$&1.55\\
			NEXT-1t&$^{136}\rm{Xe}$&\bm{$2.33 \times 10^{27}$}&$4.64 \times 10^{27}$&$2.23 \times 10^{27}$&1.99\\
			KL2Z&$^{136}\rm{Xe}$&$1.85 \times 10^{27}$&$2.78\times 10^{27}$&$1.29\times 10^{27}$&1.5\\
			JUNO (50 tons)&$^{136}\rm{Xe}$&$1.99\times 10^{28} $&$2.84\times 10^{28}$&$1.31\times 10^{28}$&1.43\\
			SNO+\uppercase\expandafter{\romannumeral2}&$^{130}\rm{Te}$&$4.3 \times 10^{26}$&$6.12\times 10^{26}$&$2.81\times 10^{26}$&1.42\\
			CUPID baseline&$^{100}\rm{Mo}$&$1.98\times 10^{27}$&$3.08\times 10^{27}$&$1.45\times 10^{27}$&1.56\\
			CUPID-1T&$^{100}\rm{Mo}$&\bm{$9.66 \times 10^{27}$}&\bm{$1.93\times 10^{28}$}&$1.09\times 10^{28}$&2.0\\
			\hline
			\hline	
		\end{tabular}		
	\end{center}
\end{table*}

The zero background condition is considered favorable. To achieve this condition for exclusion (discovery) sensitivity, the total expected background $\lambda_{b}T$ should be $<$0.69 ($<$0.0027) counts. No experiments achieve zero-background conditions of discovery sensitivity, while CDEX-1T, CDEX-10T, LEGEND-1000, and CUPID-1T reach a zero-background condition of exclusion, while NEXT-1t ($\lambda_{b}T=0.71$ when $T=10$ yr) is very close to it, as is shown in Table~\ref{ta:5yr_half-life} and Table~\ref{ta:10yr_half-life}. The ratio of ten-yr exclusion sensitivity to five-yr exclusion sensitivity is detailed in Table \ref{ta:10yr_half-life}, providing further insights. As indicated, experiments reaching zero background condition of exclusion doubles the sensitivity, while experiments with higher background level have lower ratio. This behavior aligns with the expectation that $T_{1/2}^{0\nu}\propto T$ under the zero background condition and $T_{1/2}^{0\nu}\propto \sqrt{T}$ when background levels are considerable~\cite{agostini2023toward}. Therefore, increasing the operation time is more effective for zero-background experiments.

\begin{figure}[!tbp]
	\includegraphics[width=\linewidth]{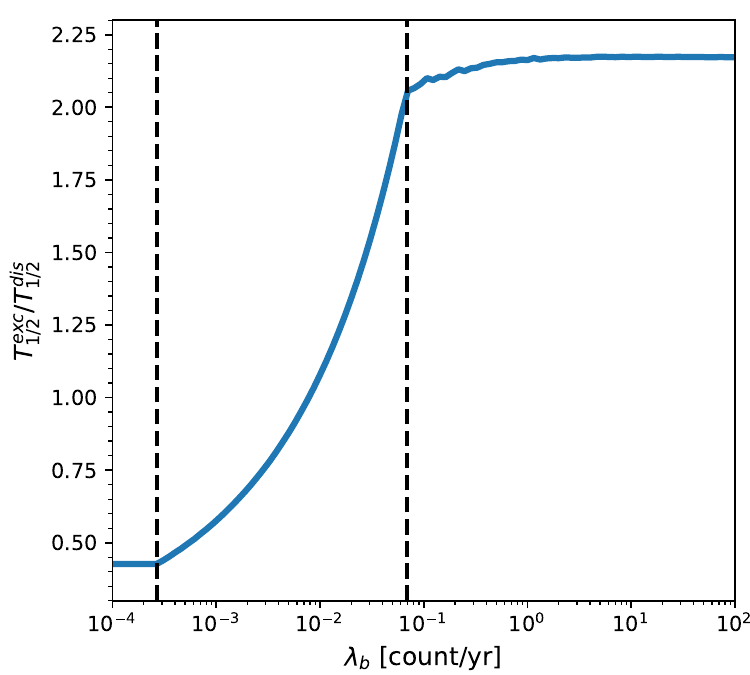}
	\caption{Ratio of exclusion sensitivity to discovery sensitivity as a function of $\lambda_{b}$ at $T=10~\rm{yr}$. Decline is observed as $\lambda_{b}$ decreases under the zero background condition of exclusion before that of discovery being reached. The two dashed lines correspond to $\lambda_{b}=0.00027$ and $\lambda_{b}=0.069$, the beginning of zero-background condition for discovery and exclusion.}
	\label{fig:halflife}
\end{figure}

In addition, while JUNO (50 tons) and CDEX-1T have similar discovery sensitivities, the exclusion sensitivity of CDEX-1T is much lower, and its exposure is only 2\% that of JUNO (1 ton vs 50 tons). This highlights the competitiveness of HPGes, which benefits from its ultrahigh technical parameter $k$ and extremely low background levels. Such an effect is demonstrated in Fig.~\ref{fig:halflife}, in which competitiveness of zero background experiments on discovery sensitivity is demonstrated.

\subsection{Ideal half-life estimation}\label{sec:4b}
For each selected experiment, an ideal scenario in which all reducible background is eliminated is considered. The upper limit of background reduction achievable in the current scenario can thus be estimated. This is particularly important because the background reduction in some LS experiments, which have relatively poor energy resolution, is constrained by the irreducible background.

For $0\nu\beta\beta$ experiments, $2\nu\beta\beta$ events whose deposit energy is within the ROI have almost the same topology as a real signal. In this subsection, with the other experimental parameters fixed, the number of $2\nu\beta\beta$ background signals for each selected experiment is provided and defined as $\lambda^{2\nu}_{b}$ ($2\nu\beta\beta$ only). In addition, the background induced by elastic scatter of electron and solar B-8 neutrinos~\cite{Solar2011} is also analyzed. Further analysis of the ideal exclusion and discovery sensitivity in the later phases is also presented.

The number of $2\nu\beta\beta$ background is calculated using Eq.~(\ref{Eq:2nu}). First, $N^{2\nu}$, the annual number of $2\nu\beta\beta$ events, is calculated. Second, $\lambda^{i}_{b}$ is calculated by multiplying $N^{2\nu}$ by the probability of events falling within the ROI and the efficiency (which is assumed to be the same as for $2\nu\beta\beta$ due to the same topology). The ROI is represented by the region $[T_{1},T_{2}]$, and $\epsilon_{ROI}$ is calculated based on the chosen ROI, as discussed in the Appendix (the ROI region is $[-2\sigma,2\sigma]$ if not mentioned). $f^{2\nu}(E)$ represents the energy distribution of $2\nu\beta\beta$ to the ground state~\cite{DBD1995}. $N(T,E,\sigma(E))$ is the normal distribution of $T$ with a mean value of $E$ and a standard deviation of $\sigma(E)$. The parameter $\eta$, used in Table~\ref{ta:LEGEND} and Table~\ref{ta:nEXO}, is included. The $\sigma(E)$ is assumed to follow the form $a\sqrt{E}$, where the parameter $a$ is calculated from $\sigma/Q_{\beta\beta}$ as given in Table~\ref{ta:LEGEND} and Table~\ref{ta:nEXO}
\begin{equation}
	\label{Eq:2nu}
	\left\{
	\begin{aligned}
		&N^{2\nu}=\rm{ln}2\frac{1}{T_{1/2}^{2\nu}}m_{iso}A\\
		&\lambda_{b}^{2\nu}=N^{2\nu}\frac{\eta}{\epsilon_{ROI}}\int_{0}^{Q_{\beta\beta}} \int_{T_{1}}^{T_{2}}N(T,E,\sigma(E))f^{2\nu}(E)dTdE
	\end{aligned}
	\right.
\end{equation}

The number of ES events per year, $\lambda_{b}^{ES}$, is estimated for each experiment at later phases through the process in~\cite{Solar2011}. $N(E_{R})$, the number of recoiled electrons (in the unit of $[\rm{keV}^{-1}\rm{yr}^{-1}(target\quad e)^{-1}]$) at the recoiled energy $T$, is calculated by Eq.~(\ref{Eq:N}): 
\begin{equation}
	\label{Eq:N}
	\begin{aligned}
		N(E_{R})=\int_{q_{min}}^{q_{max}}\Phi (E_{\nu})(P_{ee}\frac{d\sigma_{e} }{dE_{R}}+(1-P_{ee})\frac{d\sigma_{\mu ,\tau } }{dE_{R}} ) dE_{\nu}
	\end{aligned}		
\end{equation}
where $q_{max}$ is the upper limit of solar B-8 neutrino spectrum~\cite{spectrum},  $q_{min}=\frac{\sqrt{E_R^{2}+2m_{e}E_{R}}+E_{R}}{2}$ is the lower limit restricted by energy and momentum conservation laws, and $\Phi (E_{\nu})$ is the flux of solar B-8 neutrinos~\cite{spectrum}. $P_{ee}$ is the survival rate of $\nu_{e}$ when it reaches earth, calculated in~\cite{Solar2011}. $\frac{d\sigma_{e} }{dE_{R}}$ and $\frac{d\sigma_{\mu ,\tau } }{dE_{R}}$ are the cross section of $\nu_{e}$ and $\nu_{\mu,\tau}$ with electrons~\cite{Solar2011}.

Then $B$, background level per detector mass per energy width per year at $Q_{\beta\beta}$ can be calculated by Eq.~(\ref{Eq:B}):
\begin{equation}
	\label{Eq:B}
	\begin{aligned}
		B=N(Q_{\beta\beta})N_{A}\sum_{i}^{}c_{i}\frac{Z_{i}}{A_{i}}
	\end{aligned}		
\end{equation}
where $c_{i}$ is the proportion of compound i in mass, while $Z_{i}$ and $A_{i}$ are the sum of the atomic number and atomic mass of all isotopes in the compound. B in different experimental setup is summarized in Table~\ref{ta:B}. Comparing the same setups, the results in this work agree very well with those in~\cite{Solar2011}. 
\begin{table}[!htbp]
	\begin{center}
	\renewcommand{\arraystretch}{1.5}
	\caption{ES induced background level B (in the unit of $[\rm{keV}^{-1}\rm{kg}^{-1}\rm{yr}^{-1}]$) of different experimental setups that different experiments adopt. The percentage in the parens represents the enrichment rate of target isotope. The detailed recipes of LS-EnXe1 for KamLAND-Zen and LS-EnXe2 for JUNO are introduced in the Appendix. }
	\label{ta:B}
		\begin{tabular}{p{2.5cm}<{\centering}p{3.5cm}<{\centering}p{2.5cm}<{\centering}}
			\hline
			\hline
			Setup&Experiment&B ($10^{-7}$)\\
			\hline
			EnGe (90\%)&LEGEND, CDEX&1.85\\
			EnXe (90\%)&nEXO, NEXT&1.68\\
			NaXe&XLZD, PandaX-xT&1.73\\
			LS-EnXe1&KamLAND-Zen&2.39\\
			LS-EnXe2&JUNO&2.32\\
			LS-NaTe&SNO+&2.33\\
			$\rm{Li}_{2}\rm{Mn}\rm{O}_{4}$ (95\%)&CUPID&1.79\\	
			\hline
			\hline		
		\end{tabular}
	\end{center}
\end{table}
 
When B is achieved, $\lambda_{b}^{ES}$ for each experiment can be calculated through Eq.~(\ref{Eq:ES}):
\begin{equation}
	\label{Eq:ES}
	\begin{aligned}
		\lambda_{b}^{ES}=B\frac{m_{iso}A\frac{\eta}{\epsilon_{ROI}}}{f_{enr}f_{load}}\Delta E
	\end{aligned}		
\end{equation}
where $f_{load}$ is the proportion of target nuclide and its isotopes by weight. Compared with $b_{(E)}$, there is a modification of efficiency for $B$, as is shown in Eq.~(\ref{lambda_b}) and Eq.~(\ref{Eq:ES}). This is because $b_{(E)}$ is the background level being detected, while $B$ measures how many concerned events happen.

The ideal background level $\lambda^{i}_{b}=\lambda_{b}^{2\nu}+\lambda_{b}^{ES}$ for each selected experiment is presented in Table~\ref{ta:ideal_b}. Among these, all experiments except XLZD (a), PandaX-xT (a) and LS experiments reach a zero background condition of exclusion. On the other hand, none of them reaches a zero background condition of discovery when $T=10$ yr (HPGes and CCs are close). $\lambda^{i}_{b}$ values for LS experiments are of the same magnitude as $\lambda_{b}$, indicating a limitation in further background reduction. The ideal background level of $b^{i}$ or $b^{i}_{E}$ can be calculated from $\lambda^{i}_{b}$ by Eq.~(\ref{lambda_b}) or Eq.~(\ref{Eq:b_E}) and parameters from Table~\ref{ta:LEGEND} or Table~\ref{ta:nEXO}. The specific values are also demonstrated in Table~\ref{ta:ideal_b}.

\begin{table*}[!tbp]
	\begin{center}
	\renewcommand{\arraystretch}{1.5}
	\caption{$\lambda^{2\nu}_{b}$ and $\lambda_{b}^{ES}$ calculated using Eq.~(\ref{Eq:2nu}) and Eq.~(\ref{Eq:ES}) are, respectively, shown in the fourth and fifth column. The sum of them $\lambda^{i}_{b}$ is shown in the sixth column, compared to the proposed background level in the third column for experiments in Table~\ref{ta:10yr_half-life}. A value of 0 indicates that such background are negligible, and thus, the specific value is not necessary for sensitivity estimation. The last column summarizes ideal background level $b^{i}$ or $b^{i}_{E}$ of each experiment calculated from Eq.~(\ref{lambda_b}) or Eq.~(\ref{Eq:b_E}). }
	\label{ta:ideal_b}
		\begin{tabular}{p{2.3cm}<{\centering}p{1.0cm}<{\centering}p{2cm}<{\centering}p{2cm}<{\centering}p{2cm}<{\centering}p{2cm}<{\centering}p{2.5cm}<{\centering}}
			\hline
			\hline
			Experiment
			&Isotope
			& $\lambda_{b}$ $(\frac{\rm{events}}{yr})$
			& $\lambda^{2\nu}_{b}$ $(\frac{\rm{events}}{yr})$
			&$\lambda^{ES}_{b}$ $(\frac{\rm{events}}{yr})$
			&$\lambda^{i}_{b}$ $(\frac{\rm{events}}{yr})$
			& $b^{i}_{(E)}$ ($\frac{\rm{events}}{\rm{keV}\cdot\rm{kg}\cdot\rm{yr}}$)\\
			\hline
			LEGEND-1000&$^{76}\rm{Ge}$&0.044&0&0.00051&0.00051&$1.2 \times 10^{-7}$\\
			CDEX-1T&$^{76}\rm{Ge}$&0.022&0&0.00051&0.00051&$1.2 \times 10^{-7}$\\
			CDEX-10T&$^{76}\rm{Ge}$&0.044&0&0.0051&0.0051&$1.2 \times 10^{-7}$\\
			nEXO (a)&$^{136}\rm{Xe}$&0.8&0.032&0.029&0.061&$2.6 \times 10^{-7}$\\
			XLZD (a)&$^{136}\rm{Xe}$&0.17&0.00088&0.083&0.084&$1.7 \times 10^{-6}$\\
			PandaX-xT (a)&$^{136}\rm{Xe}$&0.2&0.0068&0.073&0.080&$2.1\times 10^{-6}$\\
			NEXT-1t&$^{136}\rm{Xe}$&0.071&0&0.001&0.001&$6.2\times 10^{-8}$\\
			KL2Z&$^{136}\rm{Xe}$&0.6&0.035&0.486&0.52&$8.4 \times 10^{-6}$\\
			JUNO (50 tons)&$^{136}\rm{Xe}$&67.45&20.57&26.53&47.1&$8.4 \times 10^{-6}$\\
			SNO+\uppercase\expandafter{\romannumeral2}&$^{130}\rm{Te}$&106.9&56.1&8.44&64.5&$2.2 \times 10^{-4}$\\
			CUPID baseline&$^{100}\rm{Mo}$&0.215&0&0.00054&0.00054&$2.5 \times 10^{-7}$\\
			CUPID-1T&$^{100}\rm{Mo}$&0.045&0&0.0022&0.0022&$2.5 \times 10^{-7}$\\
			\hline
			\hline 
		\end{tabular} 
	\end{center}
\end{table*}

Using the values of $\xi$ in Table~\ref{ta:LEGEND} and Table~\ref{ta:nEXO}, along with $\lambda^{i}_{b}$ from Table~\ref{ta:ideal_b}, the exclusion and discovery sensitivity for each selected experiment at the later phase are calculated according to Eq.~(\ref{Eq:halflife2}), as is listed in Table~\ref{ta:ideal_halflife}. Compared to the results in Table~\ref{ta:10yr_half-life}, the improvement for LS experiments is relatively less notable, as irreducible background already constitutes a substantial proportion of the total background. However, for non-LS experiments, a more noticeable improvement in discovery sensitivity is observed, consistent with the decrease in $T_{exc}/T_{dis}$ shown in Fig.~\ref{fig:halflife}. On the other hand, if a certain experiment has reached zero background condition of exclusion, such as LEGEND and CDEX, background reduction can not further increase exclusion sensitivity.

Since NEXT and nEXO have the potential to reduce all backgrounds except $2\nu\beta\beta$ by Ba tagging, the sensitivity when there is only a $2\nu\beta\beta$ background is also listed in Table~\ref{ta:ideal_halflife}. It is worth mentioning that NEXT is the only experiment so far that may reach zero background condition of discovery, thanks to its potential to eliminate the solar B-8 neutrino background. 

\begin{table}[!htbp]
	\begin{center}
	\renewcommand{\arraystretch}{1.5}
	\caption{Exclusion and discovery half-life sensitivity of selected experiments in an ideal situation at later stages when $T=10~\rm{yr}$. Lattice is in bold form if it is in zero background condition.}
	\label{ta:ideal_halflife}
		\begin{tabular}{p{3.2cm}<{\centering}p{0.9cm}<{\centering}p{2.2cm}<{\centering}p{2.2cm}<{\centering}}
			\hline
			\hline
			Experiment&Isotope&Exclusion $(\rm{yr})$&Discovery $(\rm{yr})$\\
			\hline
			LEGEND-1000&$^{76}\rm{Ge}$&\bm{$1.82\times10^{28}$}&$3.71\times10^{28}$\\
			CDEX-1T&$^{76}\rm{Ge}$&\bm{$1.83\times10^{28}$}&$3.72\times10^{28}$\\
			CDEX-10T&$^{76}\rm{Ge}$&\bm{$1.83\times10^{29}$}&$2.07\times10^{29}$\\
			nEXO (a)&$^{136}\rm{Xe}$&\bm{$3.53\times10^{28}$}&$1.79\times10^{28}$\\
			nEXO ($2\nu\beta\beta$ only)&$^{136}\rm{Xe}$&\bm{$3.53\times10^{28}$}&$2.25\times10^{28}$\\
			XLZD (a)&$^{136}\rm{Xe}$&$1.60\times10^{28}$&$7.71\times10^{27}$\\
			PandaX-xT (a)&$^{136}\rm{Xe}$&$9.23\times10^{27}$&$4.44\times10^{27}$\\
			NEXT-1t&$^{136}\rm{Xe}$&\bm{$4.66\times10^{27}$}&$7.75\times10^{27}$\\
			NEXT ($2\nu\beta\beta$ only)&$^{136}\rm{Xe}$&\bm{$4.66\times10^{27}$}&\bm{$1.09\times10^{28}$}\\
			KL2Z&$^{136}\rm{Xe}$&$2.96\times10^{27}$&$1.37\times10^{27}$\\
			JUNO (50 tons)&$^{136}\rm{Xe}$&$3.38\times10^{28}$&$1.56\times10^{28}$\\
			SNO+\uppercase\expandafter{\romannumeral2}&$^{130}\rm{Te}$&$7.87\times10^{26}$&$3.62\times10^{26}$\\
			CUPID baseline&$^{100}\rm{Mo}$&\bm{$4.64\times10^{27}$}&$9.31\times10^{27}$\\
			CUPID-1T&$^{100}\rm{Mo}$&\bm{$1.93\times10^{28}$}&$2.75\times10^{28}$\\
			\hline
			\hline
		\end{tabular}
	\end{center}
\end{table}

\begin{figure}[!tbp]
	\includegraphics[width=\linewidth]{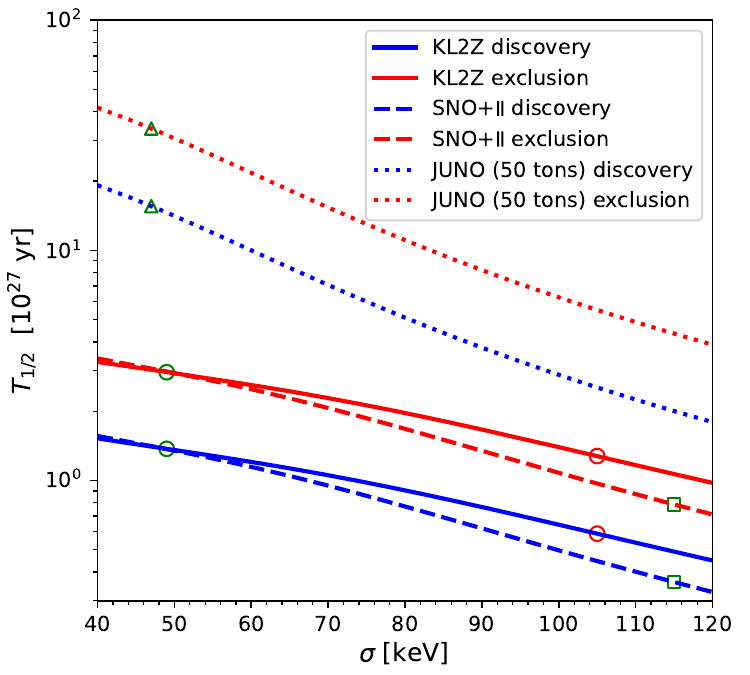}
	\caption{Relationship between the ideal $T_{1/2}^{0\nu}$ sensitivity at the later phases and $\sigma$ for KL2Z, JUNO (50 tons), and SNO+\uppercase\expandafter{\romannumeral2}, with $T = 10$ yr. Green scatters located at each curve depict the sensitivity at resolution that is depicted in Table~\ref{ta:nEXO}. Red scatters on the curve of KL2Z correspond to resolution of KLZ-800, which is a conservative projection.  }
	\label{fig:T_resolution}
\end{figure}

To further increase the sensitivity of an LS experiment, the energy resolution should be improved. The relationship between the ideal $T_{1/2}^{0\nu}$ and $\sigma$ for the three selected LS experiments is demonstrated in Fig.~\ref{fig:T_resolution}, clearly illustrating the effectiveness of upgrading energy resolution. Since an upgrade of resolution may not be as good as projection, the figure offers a comprehensive reference for future LS experiments.

\section{Estimation of mass sensitivity} \label{sec:5}
\begin{figure}[!tpb]
	\includegraphics[width=\linewidth]{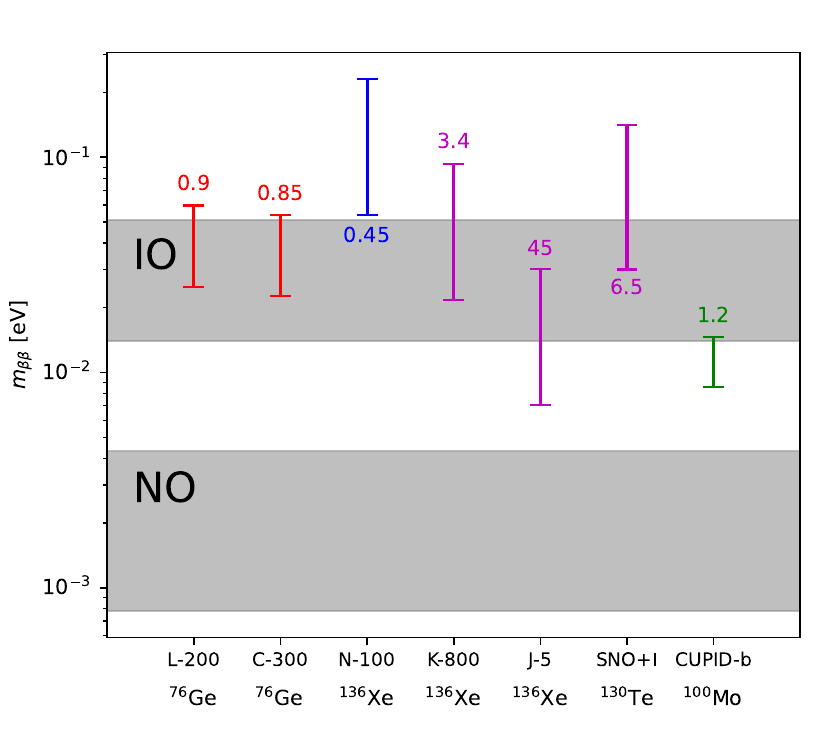}
	\caption{Exclusion $m_{\beta\beta}$ sensitivities of the selected experiments during their early phase. L-200, C-300, N-100, K-800, J-5, and CUPID-b, respectively, correspond to LEGEND-200, CDEX-300, NEXT-100, KLZ-800, JUNO (5 tons), and CUPID baseline. The sensitivities are expressed in the form of band due to $|M^{0\nu}|$ uncertainty. Nondegenerate IO ($14\sim 51$ meV) and NO bands ($0.78\sim 4.3$ meV) are demonstrated for reference. The exposure (mass of target isotope multiplies operation time) of each experiment is demonstrated in the unit of [$\rm{ton\cdot yr}$].}
	\label{fig:m_5}
\end{figure}

Calculations of the $m_{\beta\beta}$ exclusion and discovery sensitivities of the ten selected experiments at different phases as well as those in ideal conditions are detailed in this section. These calculations are based on Eq.~(\ref{Eq:halflife1}), using the values of $G^{0\nu}$ and $|M^{0\nu}|$ summarized in Table~\ref{ta:parameter} and $T_{1/2}^{0\nu}$ detailed in Table~\ref{ta:5yr_half-life}, Table~\ref{ta:10yr_half-life}, and Table~\ref{ta:ideal_halflife}. As depicted in Fig.~\ref{fig:m_5}, during early phases, LEGEND, CDEX, KamLAND-Zen, and SNO+ begin to exclude the upper bound of the IO band but cannot reach its lower bound. Conversely, JUNO and CUPID begin to exclude the lower bound of the IO band.

\begin{figure*}[!htpb]
	\includegraphics[width=\linewidth]{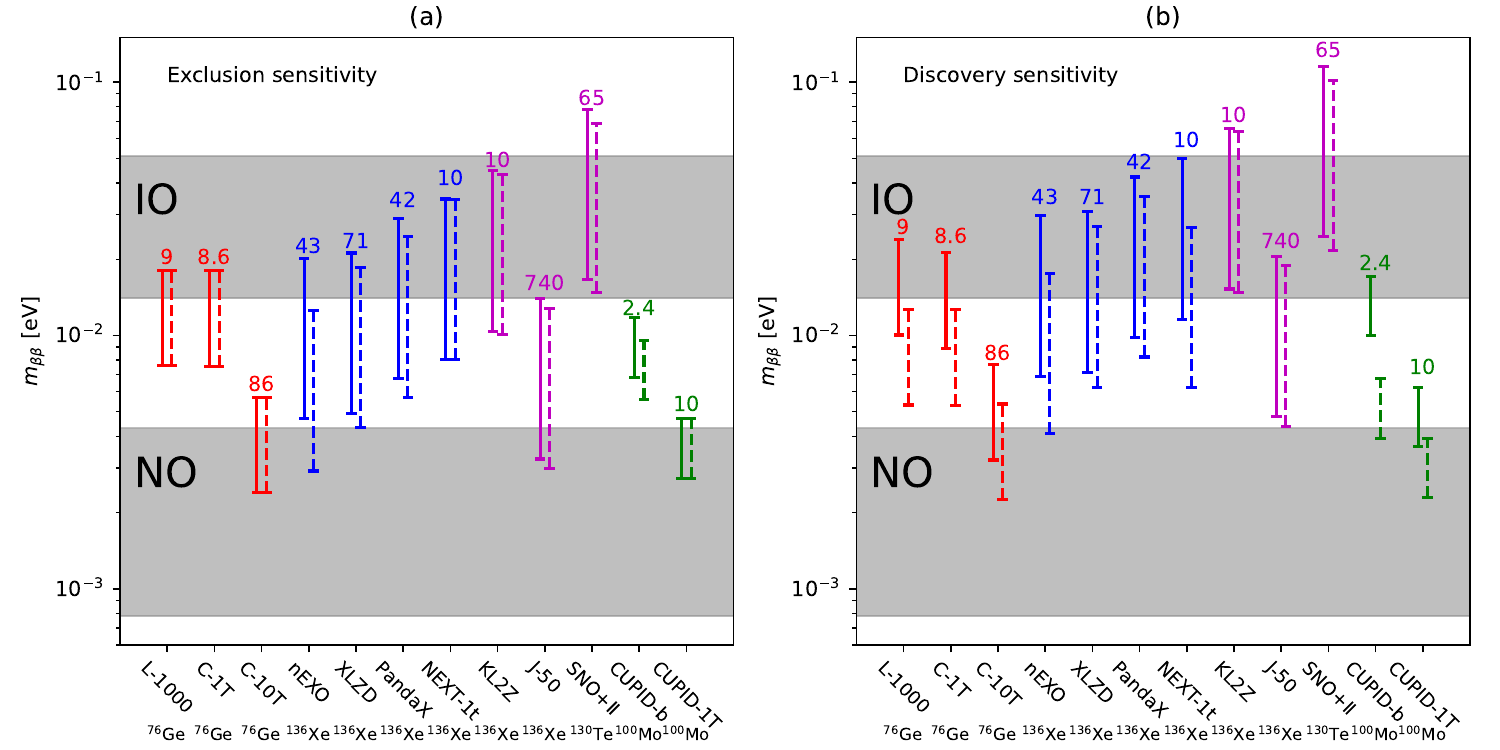}
	\caption{(a) Exclusion and (b) discovery $m_{\beta\beta}$ sensitivities of selected experiments at later phases, $T=10$ yr. L-1000, C-1T, C-10T, J-50, and CUPID-b, respectively, correspond to LEGEND-1000, CDEX-1T, CDEX-10T, JUNO (50 tons), and CUPID baseline. nEXO, XLZD, and PandaX are in their aggressive conditions. The sensitivities are represented as bands due to the uncertainty in $|M^{0\nu}|$. The nondegenerate IO ($14\sim 51$ meV) and NO bands ($0.78\sim 4.3$ meV) are included for reference. The exposure of each experiment is demonstrated at the top in the unit of [$\rm{ton\cdot yr}$].}
	\label{fig:m_10}
\end{figure*}

As illustrated in Fig.~\ref{fig:m_10}, during the later phases, CDEX-10T leads in the lower limit of the $m_{\beta\beta}$ band with CUPID-1T the second. From the exposure needed it can be found that LS experiments are relatively not efficient to increase exposure, because of their low technical parameter $k$s (shown in Table~\ref{ta:nEXO}). On the other hand, experiments in Table~\ref{ta:LEGEND} have large $k$s and therefore low background level (reaching zero background condition of exclusion). As a result, they can reach a high enough sensitivity with lower exposure. Furthermore, the estimation for CUPID baseline is also depicted in Fig.~\ref{fig:m_10}, showing that CUPID can reach similar $m_{\beta\beta}$ exclusion sensitivity with only one-fourth exposure of LEGEND and CDEX. As a result, the advantage of large $\sqrt{G^{0\nu}|M^{0\nu}|^{2}}$ is shown by CUPID both in Fig.~\ref{fig:m_5} and Fig.~\ref{fig:m_10} with its leading performance in the metric of $m_{\beta\beta}$ sensitivity. It is worth mentioning that the short uncertainty band of $^{100}\rm{Mo}$ is because of lack of shell model estimation, as has been discussed in Sec.\ref{sec:2a}.

The ideal $m_{\beta\beta}$ bands of the experiments in Fig.~\ref{fig:m_10} demonstrate a dramatic improvement in discovery sensitivity for nonLS experiments, supporting the same conclusion as discussed in Sec.~\ref{sec:4}. In ideal conditions, low-background experiments such as LEGEND, CDEX, XLZD, CUPID greatly increase discovery sensitivity. Two TPC experiments, nEXO and XLZD, can discover the NO band in ideal condition. On the other hand, because of the non-negligible contribution of irreducible signals in LS experiments, the improvement on sensitivity is limited.

\section{Summary and prospect} \label{sec:6}
As $0\nu\beta\beta$ decay experiments based on diverse technologies are being proposed or developed worldwide, it becomes essential to estimate their sensitivities and evaluate their future prospects. This study implemented multiple sensitivity metrics (exclusion and discovery, $T_{1/2}^{0\nu}$ and $m_{\beta\beta}$) to provide a comprehensive comparison of selected experiments and highlight their unique features. Additionally, sensitivities under ideal conditions, where only the irreducible background exists, are calculated to examine the limits of background reduction efforts.

First, a method for deriving $T_{1/2}^{0\nu}$ and $m_{\beta\beta}$ based on $\xi$ and $\lambda_{b}$ is adopted from~\cite{agostini2023toward,Jason2017}. The technical parameter $k=\frac{\xi}{\lambda_{b}}$ is introduced to quantify performance under fixed exposure conditions and separate the effect of different parameters on sensitivity. Notably, experimental parameters required to calculate $\xi$ and $k$ are sourced from the literature, as summarized in Table~\ref{ta:LEGEND} and Table~\ref{ta:nEXO}. Using these parameters, $T_{1/2}^{0\nu}$ and $m_{\beta\beta}$ sensitivities for ten typical experiments with promising prospects are computed using Eq.~(\ref{Eq:halflife1}) and Eq.~(\ref{Eq:halflife2}). Notably, the proposed approach is flexible, allowing calculations of sensitivity for upgraded or new experiments.

Next, the exclusion and discovery $T_{1/2}^{0\nu}$ sensitivities of the experiments are obtained, as summarized in Table~\ref{ta:5yr_half-life} and Table~\ref{ta:10yr_half-life} as well as ideal sensitivities ($2\nu\beta\beta$ background only) in Table~\ref{ta:ideal_halflife}. Multiple-phase estimation is conducted for each selected experiment. Experiments operating under zero-background conditions, such as LEGEND, CDEX, and CUPID, benefit significantly from increased operational time, excelling in discovery sensitivity. In addition, a larger improvement in discovery sensitivity for non-LS experiments and the considerable effectiveness of energy resolution upgrades for LS experiments in ideal conditions are also demonstrated.

Finally, exclusion and discovery $m_{\beta\beta}$ sensitivities for the selected experiments, as well as the ideal results, are illustrated in Fig.~\ref{fig:m_5} and Fig.~\ref{fig:m_10}. The exposure needed is also demonstrated for comparison. Based on the $T_{1/2}^{0\nu}$ sensitivity, the effects of uncertainty in $|M|^{0\nu}$ and number of $\sqrt{G^{0\nu}|M^{0\nu}|^{2}}$ are demonstrated, emphasizing the benefits of $^{100}\rm{Mo}$. Apart from CUPID, CDEX featuring a low background and JUNO featuring a large exposure lead in sensitivity.

This work puts much emphasis on irreducible background, $2\nu\beta\beta$ and $\nu-e$ ES scattering from solar B-8 neutrinos. Ideal sensitivities of selected experiments when there is only irreducible background are calculated, demonstrating the prospect of reducing background. It is found that experiments with good resolution, such as HPGes and CCs, have larger sensitivity improvement than those with relatively bad resolution, such as LSs. Larger improvement is found in discovery sensitivity. Because of bad resolution, the irreducible background of LS experiments is non-negligible, limiting the prospect of background reduction. On the other hand, the effects of improving resolution are obvious for LSs, which should be given more effort on in the proposed LS experiments.

In addition to its achievements, this study also has several limitations. (1) Only ten experiments are included, excluding other proposed experiments, which limits the completeness of this study. (2) Some experimental parameters (italicized in Table~\ref{ta:LEGEND} and Table~\ref{ta:nEXO}) are not available in the literature and are derived from declared sensitivities or directly cited from Ref.~\cite{agostini2023toward}, reducing confidence in these estimates. Moreover, for SNO+ experiments there is a proposal declaring full exploration of the IO band at SNO+\uppercase\expandafter{\romannumeral2}~\cite{SNO+_p}. However, the result is given without further analysis, so that no experimental parameter improvement about SNO+\uppercase\expandafter{\romannumeral2} is given, thus underrating its sensitivity in this work. (3) Representation on the background level is simplified to a single parameter $b_{E}$ for a large scale experiment such as LSs, where partitioning is often applied in actual experimental analysis~\cite{abe2023search,KLZComplete}. The simplification other research also uses~\cite{agostini2023toward} is rather acceptable, as is validated by comparing experimental results in~\cite{abe2023search,KLZComplete} ($2.3\times 10^{26}$ yr and $3.8\times 10^{26}$ yr ) with exclusion sensitivities calculated by the method in this work ($2.7\times 10^{26}$ yr and $4.3\times 10^{26}$ yr). (4) Background from other reducible sources like U/Th from structure material (in HPGes and CCs) actually cannot be fully eliminated in the real experimental setup. Ideal sensitivity only gives an upper limit in terms of theoretical physics for reference.

\section*{Acknowledgments} 
This work was supported by the National Key Research and Development Program of China (Grants No. 2023YFA1607110, No. 2022YFA1604701, and No. 2022YFA1605000), and the National Natural Science Foundation of China (Grants No. 12322511 and No. 12175112).

\section{DATA AVAILABILITY} \label{sec:7}
The data supporting this study's findings are available within the article.

\section*{APPENDIX: EXPERIMENTAL PARAMETERS}\label{appendix}
This Appendix gives a rather detailed discussion of the status of typical experiments analyzed in this work. The experimental parameters in Table~\ref{ta:LEGEND} and Table~\ref{ta:nEXO} can all be acquired and indicated from publications of respective collaborations. If not discussed specifically, the ROI of certain experiment is $[-2\sigma, 2\sigma]$.

Two HPGe experiments with similar design, LEGEND and CDEX, are considered in this work. The early phase of LEGEND, LEGEND-200, is already in operation and generating data. CDEX, on the other hand, though it falls behind in progress, has an edge in background control because of the higher equivalent depth of the China Jiping Underground Laboratory (CJPL) compared to that of Gran Sasso National Laboratory (LNGS). In CDEX-10T, the germanium and related material are all to be fabricated underground, providing a dramatic background cut~\cite{prospect_cjpl}. 

Details regarding LEGEND~\cite{abgrall2021legend,LEGEND200_TAUP2023} are summarized in Table~\ref{ta:LEGEND}. LEGEND-200, based on the GERDA infrastructure, utilizes 200 kg of enriched $\rm{Ge}$, while LEGEND-1000 requires new infrastructure to accommodate 1000 kg of enriched $\rm{Ge}$. The two phases, respectively, target half-life sensitivity of $10^{27}$ yr and $10^{28}$ yr. The major upgrade in LEGEND1000 compared to LEGEND200 is background reduction by a factor of 20 and exposure expansion by a factor of 5. 

The information of CDEX~\cite{CDEX300,prospect_cjpl} is also summarized in Table~\ref{ta:LEGEND}. A schedule of three phases for CDEX has been proposed, CDEX-300, CDEX-1T, and CDEX-10T. CDEX-300 and CDEX-1T are counterparts of LEGEND-200 and LEGEND-1000 with 225 kg and 1000 kg enriched Ge. CDEX-10T, with 10 t enriched Ge, has the potential to improve the lower limit of $T_{1/2}^{0\nu}$ to an unprecedented level of $10^{29}$ yr. CDEX expectedly reduces the background level by a factor of 2 compared to LEGEND at the same phase. This is because of a better shield from cosmic rays provided by CJPL. 

Similarly, details regarding nEXO, an experiment utilizing TPC technology, are outlined in Table~\ref{ta:nEXO}. The exposure for nEXO is fixed by the volume of its container. Differences in sensitivity are driven primarily by improvements in energy resolution and background reduction. The values of $\epsilon_{act}$ and $\sigma$ are sourced from Ref.~\cite{Adhikari_2022}, while $b_{E}$ values for conservative and aggressive conditions are derived from the ``baseline" and ``aggressive" scenarios outlined in Ref.~\cite{piepke2018sensitivity}, with additional estimation and calculation. The value of $\eta$ in Table~\ref{ta:nEXO} is calculated based on the ten-year exclusion sensitivity of $1.35\times 10^{28}~\rm{year}$ estimated in Ref.~\cite{Adhikari_2022}, which provides a detailed analysis of nEXO's prospects.

NEXT, another TPC with enriched Xe is summarized in Table~\ref{ta:nEXO}~\cite{NEXT_review}. Two phases for NEXT, NEXT-100 and NEXT-1t, have been proposed to increase the lower limit of $T_{1/2}^{0\nu}$ to the magnitude of $10^{27}$ yr. The background level is cited from~\cite{agostini2023toward}, while other parameters are from Ref.~\cite{NEXT_review}. The ROIs of NEXT-100 and NEXT-1t are $[-1\sigma,1.9\sigma]$ and $[-0.7\sigma,2.5\sigma]$, which is chosen for best sensitivity . NEXT and nEXO collaborations are also developing Ba tagging techniques, which have the potential to reduce all backgrounds except $2\nu\beta\beta$ signals~\cite{agostini2023toward}.

The experimental parameters of XLZD are also summarized in Table~\ref{ta:nEXO}. Two scenarios with 60 and 80 tons of natural Xe (8.9\% abundance of $^{136}\rm{Xe}$) are analyzed to estimate the sensitivity. In the optimistic XLZD scenario, values for $\epsilon_{act}$, $\eta$, $\sigma$, and the ROI region $[-\sigma,\sigma]$ can be found in Ref.~\cite{XLZD2024}. Further analysis indicates that $b_{E}$ is set to the DARWIN value (calculated in Ref.~\cite{agostini2023toward}). Due to the differing metrics used, the sensitivity estimated here is lower than that reported in Ref.~\cite{XLZD2024}.

Experimental parameters of PandaX-xT~\cite{PandaX2025}, another TPC experiment featuring natural Xe besides XLZD, are also summarized in Table~\ref{ta:nEXO}. The inner 8.4 tons liquid Xe of the total 47 tons Xe are expected to work as fiducial volume with a energy resolution of $\sigma=25~\rm{keV}$. The background levels in ``baseline'' and ``ideal'' condition are both listed and correspond to PandaX-xT (c) and PandaX-xT (a) with the same ROI region $[-\sigma,\sigma]$ as XLZD.

Three LS experiments, namely KamLAND-Zen, JUNO, and SNO+ are examined owing to their varying features. KamLAND-Zen~\cite{abe2023search,Shirai_2017} yields leading results for the lower limit of $T_{1/2}^{0\nu}$ and the upper limit of $m_{\beta\beta}$. However, further improvements are constrained by its current infrastructure. JUNO claims the largest LS detector, capable of accommodating the largest exposure (50 tons of $^{136}\rm{Xe}$ in its fiducial volume)~\cite{Zhao_2017}. However, in JUNO, $0\nu\beta\beta$ decay detection will only commence after achieving its primary goal of determining the neutrino mass ordering. SNO+~\cite{albanese2021sno+} features $^{130}\rm{Te}$ as its target isotope, different from $^{136}\rm{Xe}$ of KamLAND-Zen and JUNO.

The parameters for KamLAND-Zen summarized in Table~\ref{ta:nEXO} are primarily sourced from Ref.~\cite{agostini2023toward}, while the energy resolution for KL2Z is obtained from Ref.~\cite{Shirai_2017}. The ROI for KamLAND-Zen is defined as [0, 1.4$\sigma$]~\cite{agostini2023toward}. Given that the detector size constrains exposure, the increase in $m_{iso}$ is limited in KL2Z. There are major upgrades expected in energy resolution and background level, each by a factor higher than 2.

The summarized parameters of JUNO in Table~\ref{ta:nEXO} are all cited from a paper analyzing the potential of searching $0\nu\beta\beta$ in JUNO~\cite{Zhao_2017}. ROI of JUNO is [$-0.5$FWHM, $0.5$FWHM]. Five tons of $^{136}\rm{Xe}$ for phase1 and 50 tons of $^{136}\rm{Xe}$ for phase2 dissolved in fiducial volume is proposed. Since the estimation is relatively rough, $k$ of the two phases is the same in literature.

The parameters for SNO+ summarized in Table~\ref{ta:nEXO} are primarily obtained from a review article published in 2016~\cite{SNO2016}. In 2020, a talk~\cite{SNO2020} on the progress and prospects of KamLAND-Zen and SNO+ revealed that the ROI for SNO+ is set at $[-0.5\sigma,1.5\sigma]$, as summarized in Ref.~\cite{agostini2023toward}. Given that since 2016, there have been no additional detailed discussions found regarding the experimental parameters of SNO+, the technical parameters for SNO+\uppercase\expandafter{\romannumeral1} and SNO+\uppercase\expandafter{\romannumeral2} are assumed to be identical. These phases use 1.3 tons and 6.5 tons of $^{130}\rm{Te}$, respectively.

The components of liquid scintillators for the up-mentioned three LS experiments are necessary for calculation of $B$ in Table~\ref{ta:B}. The LS of JUNO~\cite{Zhao_2017} and SNO+~\cite{albanese2021sno+}  are mainly composed of linear alkyl-benzene (LAB) while the LS of  KamLAND-Zen~\cite{abe2023search} consists of 82\% decane and 18\% pseudocumene.

The parameters for CUPID detailed in Table~\ref{ta:LEGEND} are obtained from Ref.~\cite{CUPID2022}. The CUPID baseline is planned to be installed within the current CUORE infrastructure, using 240 kg $^{100}\rm{Mo}$. An estimated background level of $10^{-4}$ cpkky is expected at the beginning, with a future upgrade of $2\times 10^{-5}$ cpkky. During the second phase, a new cryostat is planned to accommodate 1000 kg of $^{100}\rm{Mo}$, with further background suppression to $6 \times 10^{-5}$ cpkky.
\bibliography{Sensitivity.bib}

\end{document}